\documentclass[pra,preprint]{revtex4}
\usepackage{graphicx}

\begin{document}
\title{Critical volume fraction and critical size for a cluster to nucleate}
\author{Lianfeng Long and Changming Xiao\footnote{Supported by the National
Natural Science Foundation of China under Grant No.10375024 and
Science Foundation of Hunan Educational Committee and the construct
program of the key discipline in hunan province}\footnote{
Corresponding author, Email: cmxiao@hunnu.edu.cn }} \affiliation{
Key Laboratory of Low-Dimensional Quantum Structures and Quantum
Control of Ministry of Education, and Department of Physics, Hunan
Normal University, Changsha 410081, China}

\begin{abstract}
Combined with the principle of entropy maximum and CS state
equation, the critical size for a cluster to nucleate then the
critical volume fraction of a phase transition is determined in this
paper, and our result is in good agreement with the experiments.
Furthermore, no parameter is introduced in the discussion. \noindent
\end{abstract}

\maketitle The phase behavior of a colloidal system is very complex
and it has attracted considerable attentions all these years
\cite{1,2,3,4,5,6,7,8,9,10,11,12,13,14,15,16,17,18,19,20}. It is
known that, in a colloidal system consisting of spheres, there are
excluded volumes around them \cite{1,2}. When the spheres approach
each other so closely that their excluded volumes overlap, the
spheres are packed, then a "cluster" consisting of the packed
spheres turns out. In this process, not only the free volume for
other isolated spheres but also the entropy of the system are
increased. According to the principle of entropy maximum, the large
spheres will be packed into a cluster so that the entropy of the
system is increased. If the spheres are packed into many clusters at
the same time, the packing takes place among these clusters rather
than among other spheres. As a result, the cluster will turn to be a
nucleation through the nucleation packing \cite{15}, and a phase
transition turns out. However, in the packing process, there is a
competing factor resulting entropy decreasing: when a part of large
spheres pack together, the number of isolated spheres decreases
resulting a decrease of entropy. So, there is a critical volume
fraction for large spheres to pack together. Based on the entropy
obtained by the CS state equation \cite{16,17}, the critical volume
fraction denoting the phase transition was determined, but the
results is smaller than that obtained by experiments \cite{13}. In
fact, when the spheres are packed together, their excluded volumes
are not completely overlapped, so a parameter was introduced to
describe the packing degree in Ref. \cite{14} and a better result
was obtained. However, the parameter cannot be easily determined. It
is very well if a good result can be theoretically obtained without
any parameter. So we would like to reconsider the problem. In fact,
it is known that, from a cluster to a phase transition, there is a
long way to go. If the packing process is analyzed in more detail,
we think that there is a critical size for a cluster to nucleate,
i.e., only when the cluster is larger than the critical size, can it
last for other spheres' further packing and grows larger and larger,
then a phase transition turns out. Therefore, in the theoretical
analyses, the critical size of the cluster should be included.
Obviously, this will increase the complexity of the problem. To make
the problem as simple as possible, a monodisperse suspension of
hard-spheres is studied in this paper.

Supposing that there are N hard spheres of diameter ds in a
container with a fixed volume V . As there are excluded volumes
surrounding each sphere, the entropy of this system is closely
related to the number of spheres and the free volume of the spheres
Vf, and can be written as \cite{14}
\begin{equation}
S=S(N,V_{f}),
\end{equation}
However, when \textit{k} spheres pack into a cluster, which can be
further taken as a large sphere, then the system turns to be a
binary system consisting of \textit{N}-\textit{k} hard spheres and a
large cluster, and can still be dealt with in the framework of a
binary system. In this binary colloidal system, two factors should
be noted: one is that the number of the cluster keeps to \textit{1}
while the number of the isolated spheres decreases by \textit{k};
the other factor is that the size of the cluster increases with the
increase of \textit{k}. For simplicity, the cluster consisting of
\textit{k} spheres is taken as a large sphere of diameter $d_{L}$.
Obviously, it is a function of k, i.e., $d_{L}=d_{L}(k)$. Then the
entropy of this system can be rewritten as
\begin{equation}
S=S(N_{s},V_{f},d_{L}(k)),
\end{equation}
and the variation of the entropy \emph{S} is
\begin{equation}
\delta S=\frac{\partial S}{\partial N_{S}}\delta N_{S}+\frac{\partial S}{%
\partial d_{L}}\frac{\partial d_{L}}{\partial N_{S}}\delta N_{S}
\end{equation}
where $N_{S}=N-k$ is number of the isolated spheres.

Only when the expression of entropy S is known, can Eq. (3) be
quantificationally studied. Fortunately, for a hard spheres system,
the entropy can be analytically obtained through the generalize CS
equation \cite{16,17}: for a binary mixture composed of
$N_{L}=c_{L}N$ large spheres with diameter $d_{L}$ and
$N_{S}=c_{S}N$ small spheres with diameter (where $c_{S}$,$c_{L}$
are concentrations of spheres satisfying $c_{S}+c_{L}=1$), the mean
volume $\omega$ and mean diameter \emph{d} of a sphere have the
following relation
\begin{eqnarray}
\omega =\frac{\pi }{6}\left( c_{L}d_{L}^{3}+c_{S}d_{S}^{3}\right)
=\frac{\pi }{6}d^{3}\nonumber
\end{eqnarray}
then the entropy is expressed as \cite{16,17}
\begin{equation}
S_{hs}=S_{gas}+S_{c}+S_{\eta }+S_{\sigma }
\end{equation}
where
\begin{eqnarray}
S_{gas}&=&Nk_{B}\ln(e\Omega\left(em_{L}^{c_{L}}m_{S}^{c_{S}}\frac{2\pi
k_{B}T}{h^{2}}\right) ^{3/2})\nonumber\\
S_{c}&=&-Nk_{B}\left( c_{L}\ln c_{L}+c_{S}\ln c_{S}\right)\nonumber\\
S_{\eta }&=&-Nk_{B}\left( \varsigma -1\right) \left( \varsigma
+3\right)\nonumber\\
S_{\sigma }&=&Nk_{B}(\frac{3}{2}(\varsigma ^{2}-1)y_{1}+
\frac{3}{2}(\varsigma-1)^{2}y_{2}\nonumber\\
&&-(\frac{1}{2}(\varsigma-1)(\varsigma+3)+\ln
\varsigma)(1-y_{3}))\nonumber
\end{eqnarray}
here $\varsigma=V/N$  is the average volume of each sphere, $m_{L}$
and $m_{S}$ are the mass of a large and small spheres, and the
parameters are defined as follow:
\begin{eqnarray}
\varsigma &=&\left( 1-\eta \right) ^{-1}
\nonumber\\
y_{1}&=&\frac{c_{L}c_{S}\left( d_{L}+d_{S}\right) \left(
d_{L}-d_{S}\right) ^{2}}{d^{3}}
\nonumber\\
y_{2}&=&\frac{c_{L}c_{S}d_{L}d_{S}\left(
c_{L}d_{L}^{2}+c_{S}d_{S}^{2}\right) \left( d_{L}-d_{S}\right)
^{2}}{d^{6}}
\nonumber\\
y_{3}&=&\frac{\left( c_{L}d_{L}^{2}+c_{S}d_{S}^{2}\right)
^{3}}{d^{6}}\nonumber
\end{eqnarray}
Here the constant $e=2.71828$, the free volume fraction is $1-\eta$.
Except for the cases of higher densities and(or) larger diameter
ratios of large-sphere to small-sphere, the validity of Eq.(4) is
verified by numerical results.

In the system considered in this paper, we suppose that a cluster
consisting of \emph{k} spheres can be further taken as a
large-sphere with mass  $m_{L}=km_{S}$  and diameter $d_{L}=\left(
\frac{6k}{\sqrt{2}\pi }\right) ^{1/3}d_{S}$, respectively. According
to principle of entropy maximum, the critical volume fraction and
critical size for a cluster to nucleate can then be obtained through
the follow equation:
\begin{equation}
\frac{dS}{dN_{s}}=\frac{\partial S}{\partial N_{S}}+\frac{\partial S}{%
\partial d_{L}}\frac{\partial d_{L}}{\partial N_{S}}=0
\end{equation}
where we have supposed the number of the small spheres is so large
that $N\rightarrow\infty$, then $N_{S}\rightarrow\infty$,
$c_{L}\rightarrow 0$, $c_{S}\rightarrow 1$, so
\begin{eqnarray}
\frac{\partial S}{\partial N_{S}}&=&k_{B}( \ln \frac{V}{N_{S}+1}+\frac{3}{%
2}\ln \frac{2\pi em_{s}k_{B}T}{h^{2}}) \nonumber\\&&-k_{B}\left(
\varsigma
^{2}+2\varsigma -3\right) -\frac{\pi \left( N_{S}+1\right) }{3V}%
k_{B}d_{S}^{3}\left( \varsigma ^{3}+\varsigma ^{2}\right)
\end{eqnarray}
By the same way we also get:
\begin{eqnarray}
\frac{\partial S}{\partial d_{L}} &=&\frac{\partial S_{gas}}{\partial d_{L}}+%
\frac{\partial S_{c}}{\partial d_{L}}+\frac{\partial S_{\eta
}}{\partial
d_{L}}+\frac{\partial S_{\sigma }}{\partial d_{L}} \nonumber\\
&=&\frac{3\sqrt{2}k_{B}}{4d_{S}k}\left( \frac{6k}{\sqrt{2}\pi
}\right) ^{2/3}+\frac{\pi \left( N_{S}+1\right)
k_{B}}{V}d_{L}^{2}\left( \varsigma
^{3}+\varsigma ^{2}\right)  \nonumber\\
&&+k_{B}( \left( \varsigma ^{2}+2\varsigma -\ln \varsigma -3\right)
\frac{\partial y_{1}}{\partial d_{L}}+\left( \varsigma
^{2}-\varsigma -\ln \varsigma \right) \frac{\partial y_{2}}{\partial
d_{L}})
\end{eqnarray}
\begin{eqnarray}
\frac{\partial y_{1}}{\partial
d_{L}}&=&\frac{c_{S}^{2}d_{S}^{3}\left( 3d_{L}+d_{S}\right) \left(
d_{L}-d_{S}\right) }{d^{6}}
\\
\frac{\partial y_{2}}{\partial
d_{L}}&=&\frac{c_{S}^{3}d_{S}^{6}\left( 3d_{L}-d_{S}\right) \left(
d_{L}-d_{S}\right) }{d^{9}}
\\
\frac{\partial d_{L}}{\partial N_{S}}&=&-\frac{d_{S}}{3}\left( \frac{6k}{\sqrt{%
2}\pi }\right) ^{-2/3}
\end{eqnarray}
Now we see that the relation between the volume fraction   and the
critical size of cluster k is described Eq.(5). Just like Ref.
\cite{14}, we suppose that the diameter and mass of a small sphere
are $d_{S}=6.9\times10^{-8}$m, and $m_{S}=3.2\times10^{-24}$Kg, and
the temperature $T=300$K. Then the relation between $\eta$ and
\emph{k} is numerically determined, and the results are shown in
Fig. 1 and Table I. From Fig.1, it is evident that the critical size
for a cluster to nucleate decreases dramatically with the increase
of the volume fraction, and it almost tends to zero when
$\eta=0.50$, so the critical volume fraction for a cluster to
nucleate is no larger than 0.5. In addition, from table I, it is
known that, when $\eta=0.492$, the critical size for a cluster to
nucleate is 1968; however, when $\eta=0.495$, the critical size is
77; when $\eta=0.50$, the critical size is only 9. In fact, we know
that the possibility for 1968 spheres to pack into a cluster with
$\eta=0.492$ is less than that for 77 spheres to pack into a cluster
with $\eta=0.495$. So our result is reasonable. Combined with Fig. 1
and Table I, we get that the nucleation packing will take place when
$0.495<\eta<0.50$. On the other hand, in the system of monodisperse
suspension of hard-spheres, the experimental results show that the
critical volume fraction denoting phase transition from colloidal
fluid to a phase of fluid and crystal in coexistence is 0.495[6].
Obviously, the phase transition takes place through the cluster's
nucleation packing. Therefore, our result is in good agreement with
the experiment's, and the reasonableness of the idea about the
critical size for a cluster to nucleate is then verified.

However, there are two factors should be mentioned in our
discussions: One is the cluster is cursorily treated as a large
sphere. In fact, a cluster is not a large sphere. The other factor
is about the validity of CS equation for a very dense system such as
$\eta=0.495$. Obviously, these two factors will affect our result in
a way. We believe that our result can be slightly modified if a more
accurate equation of state is used to deal with the nucleation
packing. But the reasonableness of our model and the qualitative of
our result cannot be changed by them completely. On the other hand,
the reasonableness of our model will be further verified if a system
with a high size asymmetry is considered. Because in this system the
critical volume fraction will be very small in which the validity of
CS equation is certain. So, for further studies, the system with a
high size asymmetry will be considered. From above discussion, we
have shown that the critical size of the cluster is very important
for nucleation packing. Taking the critical size into consideration,
the critical volume fraction denoting a phase transition can be
determined without any parameter. Compared with the model discussed
in Ref. \cite{14}, in which a parameter was introduced to modify the
packing process, the ideal about the critical size for cluster to
nucleate shows more information about the nucleation packing.

\begin{center}
\begin{tabular}
 {|c|c|c|c|c|c|c|c|c|c|}
  \hline
  Volume fraction $\eta$ & 0.492 & 0.493 & 0.494 & 0.495 & 0.496 & 0.497 & 0.498 & 0.499 & 0.50 \\
  \hline
 Cluster size ~~~\emph{k} & 1968 & 430 & 160 & 77 & 43 & 27 & 18 & 12 & 9 \\
  \hline

\end{tabular}
\end{center}
Table I The critical volume fraction and the corresponding critical
size for cluster to nucleate.
\begin{figure}[htp]
\begin{center}
\includegraphics[width=12cm,height=10cm]{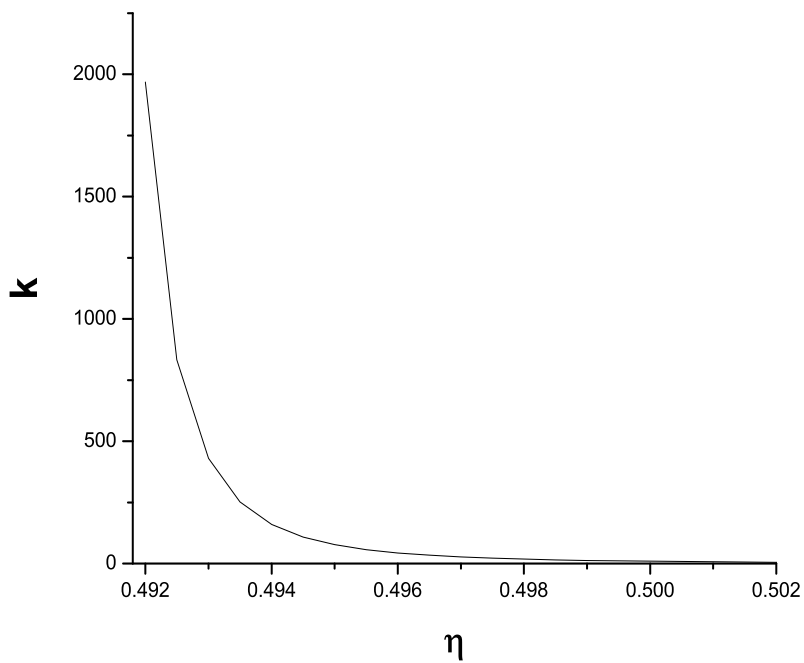}
\caption{The relation between the critical volume fraction $\eta$
and the critical size of the cluster \emph{k}.}
\end{center}
\end{figure}
\end{document}